\documentclass[%
 reprint,
 amsmath,amssymb,
 aps,
nofootinbib,
]{revtex4-2}

\usepackage{graphicx}
\usepackage{dcolumn}
\usepackage{bm}
\usepackage{hyperref}

\begin{document}

\preprint{2511.06640}

\title{Starobinsky Inflation and the Latest CMB Data:\\A Subtle Tension?}

\author{J. Bezerra-Sobrinho}
\email{jeremias.bezerra.100@ufrn.edu.br}
\affiliation{%
Departamento de Física, Universidade Federal do Rio Grande do Norte \\ Campus Universitário, s/n - Lagoa Nova, CEP 59072-970, Natal, Rio Grande do Norte, Brazil
}%

\author{L. G. Medeiros}
\email{leo.medeiros@ufrn.br}
\affiliation{%
Escola de Ciências e Tecnologia, Universidade Federal do Rio Grande do Norte \\ Campus Universitário, s/n - Lagoa Nova, CEP 59072-970, Natal, Rio Grande do Norte, Brazil
}%

\date{\today}

\begin{abstract}
We analyze the Starobinsky inflation model and the impact of curvature corrections, particularly a cubic $R^3$ term, to assess their behavior in light of the latest observational results from the Atacama Cosmology Telescope (ACT). With the recent sixth data release (DR6), the scalar spectral index was measured to be $n_s=0.9743 \pm 0.0034$, which appears to exclude the pure Starobinsky model at approximately the $2\sigma$ level. In this paper, we implement the Starobinsky inflationary potential directly into the CLASS code, without relying on the slow-roll approximation, and we constrain the number of e-folds of inflation $N_k$ using a theoretically motivated range derived from reheating considerations and standard couplings between matter fields and gravity. We show that it is still possible to identify a significant region of parameter space where the Starobinsky model remains highly consistent with the latest observational data. While the pure Starobinsky model remains a compelling candidate for cosmic inflation, we explore how including a cubic  $R^3$ term can shift its predictions to better align with the Planck and ACT measurements. 
\end{abstract}

\maketitle


\section{\label{sec:I}Introduction}

One of the key aspects of cosmology is understanding the origin of the anisotropies observed in the cosmic microwave background (CMB) and the large-scale structure of the universe. Over the past decade, these anisotropies have been measured with remarkable precision, allowing us to determine the six parameters of the standard $\Lambda$CDM model with remarkable accuracy, while also placing stringent constraints on possible extensions of the model. In recent years, precise measurements of the CMB power spectra from various observations, including space observatories such as Planck and WMAP, and ground-based experiments like the South Pole Telescope (SPT) and the Atacama Cosmology Telescope (ACT), have yielded increasingly accurate and detailed maps of the microwave sky.

Regarding the anisotropies present in the CMB, inflation provides a natural mechanism for amplifying quantum vacuum fluctuations~\cite{Guth1981, Linde1983, Mukhanov2005}. Among the many inflationary models proposed~\cite{Martin2014, Myrzakulov2015, Elizalde2017}, the Starobinsky model~\cite{Starobinsky1980}, based on $R+R^2$ gravity, has long been regarded as one of the most successful frameworks for explaining the early universe's accelerated expansion. Its predictions for the scalar spectral index $n_s$ and tensor-to-scalar ratio $r$ are in excellent agreement with Planck and BICEP/Keck results~\cite{Planck2018X, BICEPKeck2021}, making it a leading candidate for describing cosmic inflation. 

However, recent measurements from the ACT~\cite{ACTDR6a, ACTDR6b} reveal subtle deviations that could hint at new physics beyond the simplest Starobinsky model. When combining the high-resolution ACT DR6 data with the large-scale CMB observations from Planck and the second data release of BAO from the Dark Energy Spectroscopic Instrument (DESI DR2)~\cite{DESIDR2}, the results appear to exclude the Starobinsky inflationary model at approximately the $2\sigma$ level. This emerging exclusion is consistent with several recent analyses in the literature~\cite{Zharov2025, Gialamas2025JCAP, Addazi2025PLB, McDonough2025, Ellis2025, Aoki2025, Odintsov2025, Cheong2025, Linde2025, Yuennan2025, Kallosh2025, Drees2025, Gialamas2025PRD, Kouniatalis}, which have identified similar discrepancies when confronting Starobinsky-like inflation with CMB recent measurements.

It is important to note that this conclusion is primarily from a rather simplified approach, in which the extreme values of $n_s$ inferred within the $\Lambda$CDM model are directly used to estimate the corresponding range of the number of e-folds $N_k$. In this approach, an inherently inflationary parameter is constrained using a model that does not fully incorporate the dynamics of inflation. Furthermore, the number of e-folds cannot take arbitrary values; it is theoretically constrained by considerations such as the reheating history of the Universe, which provides a more physically motivated range for $N_k$.

In this paper, we perform a more rigorous analysis by implementing the inflationary potential directly into Boltzmann codes such as CLASS~\cite{Lesgourgues2011CLASS, Blas2011JCAP}. This approach allows us to constrain the number of e-folds of inflation, $N_k$, within a theoretically motivated range consistent with reheating considerations and standard matter–gravity couplings. At the same time, it enables us to identify a broad region of parameter space where the Starobinsky model remains fully compatible with the latest observational data from Planck and ACT.

Although the pure Starobinsky model remains a compelling candidate for cosmic inflation, we also investigate the effects of introducing an additional term into the system. Motivated by this consideration, we analyze the impact that a higher-order curvature correction produces on Starobinsky inflation. Before proceeding, we briefly outline the context in which higher-order corrections arise.

From a theoretical standpoint, General Relativity (GR) may be understood as a consequence of Lovelock’s theorem \cite{Lovelock1971}. Formulated by David Lovelock, this theorem asserts that ``\textit{the only second-order differential equation that can be derived from an action depending solely on the metric in four dimensions is Einstein's equation}" \cite{Clifton2012}. More precisely, Lovelock’s theorem relies on four fundamental assumptions: \emph{(a)} the action is defined over a four-dimensional integration domain; \emph{(b)} the metric constitutes the unique dynamical field; \emph{(c)} the resulting field equations are invariant under diffeomorphism transformations; and \emph{(d)} the field equations are at most second order in derivatives. Deviations from GR can be constructed by relaxing any of these underlying assumptions. For example, if the condition that the metric is the only fundamental field is abandoned and an additional scalar degree of freedom is introduced, one arrives at Horndeski theories \cite{Horndeski1974}. Alternatively, allowing the gravitational field equations to contain derivatives of order higher than two, while keeping the remaining assumptions intact, leads to the class of higher-order gravity theories \cite{Cuzinatto2008,Cuzinatto2016}. The $R+R^2$ model and quadratic gravity provide simple examples of higher-order theories.

Higher-order gravity theories are built from curvature invariants constructed out of the Riemann tensor and covariant derivatives. A systematic way to classify the individual terms in the higher-order gravitational action is through their mass dimension. Working in natural units, the covariant derivative $\nabla_{\mu}$ carries mass dimension one, whereas the Riemann tensor $R_{\mu\nu\alpha\beta}$ has mass dimension two. As a simple example, the Ricci scalar appearing in the Einstein–Hilbert action corresponds to an operator of mass dimension two. At mass dimension four, there exist four independent curvature invariants; however, only two of them contribute to the field equations:
\begin{equation}
R^2 \text{ \ \ and \ \ } R^{\mu\nu}R_{\mu\nu}.
\end{equation}
At mass dimension six, on the other hand, there are eight relevant invariants, which can be chosen as \cite{Decanini2007, Fulling1992}:
\begin{equation}
\begin{aligned}
& R^3, \quad
R\, R_{\mu\nu} R^{\mu\nu}, \quad
R\, R_{\alpha\mu\beta\nu} R^{\alpha\mu\beta\nu}, \quad \\
&R_{\alpha\varphi} R^{\rho}{}_{\mu\beta\nu} R^{\alpha\varphi\mu\beta\nu}, \quad
R_{\alpha\mu\beta\nu}R_{\text{ }\rho\text{ \ \ }\sigma}^{\alpha\text{ \ }\beta}R^{\mu\rho\nu\sigma},
\\
&R_{\alpha\mu\beta\nu}R_{\text{ \ \ \ }\rho\sigma}^{\alpha\mu}%
R^{\text{ \ \ }\rho\sigma\beta\nu},
\quad R \Box R, \text{ \ \ \ and \ \ \ } R_{\mu\nu} \Box R^{\mu\nu}.
\end{aligned}
\end{equation}
Within the effective field theory framework, this classification provides a natural interpretation of higher-order operators as higher-energy corrections to GR. In this context, the leading corrections to the Einstein–Hilbert action arise at mass dimension four, while subleading contributions correspond to operators of mass dimension six, and so on.

In order to study a simple deformation of Starobinsky inflation, we introduce an additional term to the $R+R^2$ model. Within a bottom-up effective field theory approach, the most natural correction at mass dimension four is the $R^{\mu\nu}R_{\mu\nu}$ term. However, this operator is known to suffer from pathologies associated with ghost degrees of freedom \cite{DeFelice2023}. The next relevant corrections appear at mass dimension six, consisting of eight independent operators, among which the $R^3$ term is the simplest. Indeed, by including only this term, the theory retains a single additional scalar degree of freedom, remains free of ghosts, and continues to fall within the $f(R)$ framework. Beyond its simplicity, $R^3$ inflation has the practical advantage that the model smoothly reduces to pure Starobinsky inflation in the limit $\alpha_0 = 0$. This allows us to test both the extended theory and its Starobinsky limit within a unified framework.

The purpose of introducing the $R^3$ term is to investigate how a minimal deformation of the Starobinsky model affects its level of agreement with CMB observations. We examine the $R^3$ model behavior first within a simplified approach, in which $\Lambda$CDM data are used to infer the corresponding range of $N_k$, and then in a more complete treatment, where the inflationary potential is implemented directly into the Boltzmann code CLASS.

Before starting, let us discuss a few points about the notations in this paper. We shall work with natural units such that $\hbar=c=1$, and the reduced Planck mass is $M_{P}=\left(8\pi G\right)^{-1/2}\approx2.4\times10^{18}GeV$. The signature of the metric is followed by $\left(-,+,+,+\right)$ and the background is spatially flat Friedmann-Lemaître-Robertson-Walker (FLRW).

The paper is organized as follows. In Section~\ref{sec:II}, we provide a brief theoretical background on the Starobinsky inflationary model and its extension, including the $R^3$ term. Section~\ref{sec:III} presents the construction of the range for the number of e-folds by considering the reheating history of the Universe in a general way, which is then applied to both the Starobinsky and $R^3$ models, constraining them through a simplified $\Lambda$CDM-based approach. In Section~\ref{sec:IV}, we perform a complete statistical analysis by implementing both models directly into the CLASS Boltzmann code. Finally, Section~\ref{sec:V} is devoted to the conclusions.

\section{\label{sec:II}Cosmic Inflation}

On large scales ($\gtrsim 100 Mpc$), we can consider the universe homogeneous and isotropic. Moreover, in the case of a spatially flat universe, the line element describing the evolution of a comoving reference frame is given by
\begin{equation}
\label{eq:FLRW_Metric}
\begin{aligned}
ds^2 = -dt^2 +a^2(t)[dx^2+dy^2+dz^2], 
\end{aligned}
\end{equation}
where $a(t)$ is the scale factor. Within the inflationary framework described by a canonical scalar field $\chi$, the metric field equations reduce to two independent relations, namely the Friedmann equations:
\begin{align}
H^2 &= \frac{1}{2}\left[\frac{1}{2}\dot\chi^2+V(\chi)\right], \label{eq:First_FriedmannEquation}\\
\dot H &= -\frac{3}{2}\left[\frac{1}{2}\dot \chi^2\right],  \label{eq:Seccond_FriedmannEquation}
\end{align}
where the dot represents the time derivative, $H \equiv \dot a/a$ is the Hubble parameter, and $V(\chi)$ is the inflationary potential. In addition to these equations, the $\chi$-field equation is given by
\begin{equation}
\label{eq:Chi_FieldEquation}
\begin{aligned}
\ddot \chi+3H\dot \chi + V'(\chi)=0.
\end{aligned}
\end{equation}
The potential $V(\chi)$ and its derivative $V'(\chi)$ are determined by the specific inflationary model. To recover the usual notation and dimensions of the scalar field and its potential, one must perform
\begin{equation}
\label{eq:usual_notation}
\begin{aligned}
\chi=\sqrt{\frac{2}{3}}\frac{\phi}{M_P}
\quad
\text {and}
\quad
\bar{V}(\phi)=\frac{3M^2_P}{2}V(\chi).
\end{aligned}
\end{equation}

The next step is to investigate the solution of these equations in an inflationary regime (quasi-exponential expansion). This regime is reached if $H$ is approximately constant, that is
\begin{equation}
\label{eq:skow-roll_condition}
\begin{aligned}
\frac{|\dot H|}{H^2} \leq 1.
\end{aligned}
\end{equation}
Furthermore, for cosmic inflation to be successful, it must last long enough to produce a sufficient number of e-folds to solve the horizon and flatness problem~\cite{Mukhanov2005}. These conditions can be achieved by requiring the slow-roll parameters
\begin{align}
\epsilon &\equiv -\frac{\dot H}{H^2}, \label{eq:epsilon_definition}\\
\eta &\equiv -\frac{1}{H}\frac{\dot \epsilon}{\epsilon}, \label{eq:eta_definition}
\end{align}
to be small.
In addition, the duration of inflation, quantified by the number of e-folds $N_k$ occurring between the time a given mode $k$ crosses the horizon and the end of inflation, is given by:
\begin{equation}
\label{eq:Number-efolds_definition}
\begin{aligned}
N_k \equiv \ln \left(\frac{a_{end}}{a}\right)=\int^{t_{end}}_{t}H_kdt.
\end{aligned}
\end{equation}
If we use the slow-roll approximation $(\epsilon,\eta\ll1)$ in the Eqs. (\ref{eq:epsilon_definition}) and (\ref{eq:eta_definition}), we get
\begin{align}
\epsilon & \approx \frac{1}{3}\left[\frac{V'(\chi)}{V(\chi)}\right]^2, \label{eq:epsilon_slow-roll}\\
\eta & \approx \frac{4}{3}\left[\frac{V''(\chi)}{V(\chi)}-\left(\frac{V'(\chi)}{V(\chi)}\right)\right]^2, \label{eq:eta_slow-roll}\\
N_k & \approx \frac{3}{2}\int^{\chi}_{\chi_{end}}\frac{V(\chi)}{V'(\chi)}d\chi,  \label{eq:Number-efolds_slow-roll}
\end{align}
where $V(\chi)$ is the potential and $V'(\chi)$ its derivative. Thus, given a specific model, we can derive the slow-roll expressions for each case.

Inflation not only solves the flatness and horizon problems, but also provides the mechanism for generating cosmological perturbations, which seed both the cosmic microwave background (CMB) anisotropies and the large-scale structure (LSS) of the Universe. At leading order, these perturbations can be decomposed into three distinct classes: scalar, vector, and tensor modes. Observations of the CMB have provided unprecedented detail about the scalar power spectrum, which is usually expressed as~\cite{Planck2018X}
\begin{equation}
\label{eq:Scalar_Power_Spectrum}
\begin{aligned}
P(k)=A_s\left(\frac{k}{k_0}\right)^{n_s-1},
\end{aligned}
\end{equation}
where $A_s$ is the amplitude of the scalar fluctuation at reference scale $k_0$ and $n_s$ describes the scale-dependence of the perturbations.

The vector perturbations decay rapidly due to the Universe's expansion and, consequently, their contribution to the primordial power spectrum is usually negligible~\cite{Piattella2018}. 

The tensor perturbations, which correspond to primordial gravitational waves, have not yet been detected. Nevertheless, their statistical properties can be described by a power spectrum, which characterizes the distribution of their amplitudes as a function of scale. Analogous to the scalar case, the tensor power spectrum is often modeled using a power-law parameterization~\cite{Planck2018X}:
\begin{equation}
\label{eq:Tensor_Power_Spectrum}
\begin{aligned}
P(k)=A_t\left(\frac{k}{k_0}\right)^{n_t},
\end{aligned}
\end{equation}
where $A_t$ is the reference amplitude of the tensor fluctuations and $n_t$ describes the scale-dependence of the perturbations. From the point of view of current observations, the anisotropies in the CMB constrain the scalar quantities and establish an upper limit for the tensor ones~\cite{Planck2018X, BICEPKeck2021}. In this context, it is convenient to describe the upper constraint with the tensor-scalar ratio $r\equiv A_t/A_s$.

In the slow-roll inflation framework, the power spectrum generated during inflation can be directly related to the initial conditions that give rise to the anisotropies observed in the CMB~\cite{Baumann2018, Rodrigues2022, Bezerra2023}:
\begin{align}
A_s & = \frac{9}{16\pi^2M^2_P}\frac{V^3_k}{V'^2_k}, \label{eq:As_slow-roll}\\
n_s & = 1+\eta-2\epsilon, \label{eq:ns_slow-roll}\\
r & = 16\epsilon,  \label{eq:r_slow-roll}
\end{align}
where we note that, in the slow-roll regime (at leading order), only three out of the four observational parameters ($A_s$, $n_s$, $A_t$, and $n_t$) are independent.

Having established the general slow-roll framework and its connection to the observable quantities, we now focus on specific inflationary potentials that can be confronted with current data. In the following subsections, we explicitly compute the corresponding observables for the Starobinsky and $R^3$ models.

\subsection{Starobinsky model}

One of the earliest models of cosmic inflation was proposed by Alexei Starobinsky in 1980 Ref.~\cite{Starobinsky1980}. It naturally arises from extending the Einstein–Hilbert action to include a quadratic curvature term, $R^2$.
\begin{equation}
\label{eq:Starobinsky_Action}
\begin{aligned}
S(g_{\mu \nu})=\frac{M^2_P}{2}\int d^4x\sqrt{-g}\left[R+\frac{1}{2\kappa_0}R^2\right],
\end{aligned}
\end{equation}
where $\kappa_0$ has squared mass units. Although it was among the first inflationary models proposed, it remains one of the most successful, providing an excellent fit to Planck and BICEP3/Keck data~\cite{Planck2018X, BICEPKeck2021}. The action \eqref{eq:Starobinsky_Action} can be conveniently rewritten as a scalar-tensor theory in the Einstein frame, in which we have a description through an auxiliary scalar field $\chi$ minimally coupled with general relativity~\cite{Mishra2020}
\begin{equation}
\label{eq:Starobinsky_Action_Chi}
\begin{aligned}
\bar S(\bar g_{\mu \nu}, \chi)=\frac{M^2_P}{2}\int d^4x\sqrt{-\bar g}\left[\bar R-3 \left(\frac{1}{2}\bar\partial^\rho\chi\bar\partial_\rho\chi +V(\chi) \right)  \right],
\end{aligned}
\end{equation}
whose associated potential is
\begin{equation}
\label{eq:Starobinsky_potential}
\begin{aligned}
V(\chi)={\frac{\kappa_0}{6}}\left(1-e^{-\chi}\right)^2.
\end{aligned}
\end{equation}
The barred quantities are defined with respect to the conformal metric $\bar g_{\mu \nu}=e^\chi g_{\mu \nu}$ and the dimensionless field $\chi$ is defined as~\cite{Rodrigues2022}

\begin{equation}
\label{eq:Chi_function_starobinsky}
\begin{aligned}
e^\chi=1+\frac{R}{\kappa_0}.
\end{aligned}
\end{equation}

The next step is to obtain $V(\chi)$, $V'(\chi)$, and $V''(\chi)$ at leading order in the slow-roll approximation, by defining the slow-roll parameter $\delta=e^{-\chi_k}$. We write the potential \eqref{eq:Starobinsky_potential} and its derivatives as
\begin{align}
V(\chi) & \approx \frac{\kappa_0}{6}\left(1-2\delta\right), \label{eq:Vsta_slow-roll}\\
V'(\chi) & \approx \frac{\kappa_0}{3}\delta, \label{eq:dVsta_slow-roll}\\
V''(\chi) & \approx -\frac{\kappa_0}{3}\delta.  \label{eq:ddVsta_slow-roll}
\end{align}
Using these results in Eqs. \eqref{eq:epsilon_slow-roll} and \eqref{eq:eta_slow-roll}, we obtain, at leading order in the slow-roll approximation
\begin{align}
\epsilon & \approx \frac{4}{3}\delta^2, \label{eq:epsilon_starobinsky}\\
\eta & \approx -\frac{8}{3}\delta.  \label{eq:eta_starobinsky}
\end{align}
These expressions clearly show that for $\delta \ll1$ we have $\epsilon \ll1$ and $\eta \ll1$, indicating that the Universe is in a (quasi) de Sitter expansion regime. The next step is to compute the number of e-folds $N_k$. Using Eqs. \eqref{eq:Vsta_slow-roll} and \eqref{eq:dVsta_slow-roll} in \eqref{eq:Number-efolds_slow-roll}, we obtain
\begin{equation}
\label{eq:Number-efolds_starobinsky}
\begin{aligned}
N_k \approx \frac{3}{4\delta}.
\end{aligned}
\end{equation}
Finally, it is convenient to express the slow-roll parameters $\epsilon$ and $\eta$ in terms of the number of e-folds. By substituting Eq. \eqref{eq:Number-efolds_starobinsky} into Eqs. \eqref{eq:epsilon_starobinsky} and \eqref{eq:eta_starobinsky}, we get
\begin{align}
\epsilon & \approx \frac{3}{4N^2_k}, \label{eq:epsilon_Nk_starobinsky}\\
\eta & \approx -\frac{2}{N_k}.  \label{eq:eta_Nk_starobinsky}
\end{align}
With these results, we can now relate the theoretical quantities to the observational parameters using Eqs.~\eqref{eq:As_slow-roll}, \eqref{eq:ns_slow-roll}, and \eqref{eq:r_slow-roll}:
\begin{align}
\kappa_0 & \approx 72\pi^2A_s\left(\frac{M_P}{N_k}\right)^2, \label{eq:kappa_starobinsky}\\
n_s & \approx 1-\frac{2}{N_k}, \label{eq:ns_starobinsky}\\
r & \approx \frac{12}{N^2_k}.  \label{eq:r_starobinsky}
\end{align}
The determination of the scalar spectral index has a direct impact on the estimation of the number of e-folds of inflation $N_k$ as can be seen from Eq. \eqref{eq:ns_starobinsky}. 

In the next section, we constrain $N_k$ within its theoretically motivated range. While the Starobinsky model remains fully consistent with the Planck measurement of the scalar spectral index, $n_s = 0.9649 \pm 0.0042$~\cite{Planck2018X}, the most recent ACT DR6 data, reporting $n_s = 0.9743 \pm 0.0034$~\cite{ACTDR6a}, indicate a deviation of nearly $2\sigma$ from the model’s prediction.

\subsection{High-order curvature ($R^3$) model}

A generalization of Starobinsky inflation obtained by including the cubic term $R^3$ in the gravitational action can be written as
\begin{equation}
\label{eq:R3_Action}
\begin{aligned}
S(g_{\mu \nu})=\frac{M^2_P}{2}\int d^4x\sqrt{-g}\left[R+\frac{1}{2\kappa_0}R^2+\frac{\alpha_0}{3\kappa^2_0}R^3\right],
\end{aligned}
\end{equation}
where $\kappa_0$ has dimensions of mass squared and $\alpha_0$ is a dimensionless parameter.

The theoretical framework and dynamical aspects of the $R^3$ model, such as stability conditions, field equations, and the detailed modeling of the inflationary phase, have been discussed in previous works \cite{Rodrigues2022, Cheong2020, Ivanov2022, Addazi2025}. Therefore, in the present paper, we provide only a concise overview of the model, restricted to the key equations necessary for the Bayesian inference and comparison with current cosmological observations.

The action \eqref{eq:R3_Action} can be conveniently rewritten as a scalar-tensor theory in the Einstein frame, in which we have a description through an auxiliary scalar field $\chi$ minimally coupled with gravitation. By following the steps presented in Ref.~\cite{Rodrigues2022}, we write
\begin{equation}
\label{eq:R3_Action_Chi}
\begin{aligned}
\bar S(\bar g_{\mu \nu}, \chi)=\frac{M^2_P}{2}\int d^4x\sqrt{-\bar g}\left[\bar R-3 \left(\frac{1}{2}\bar\partial^\rho\chi\bar\partial_\rho\chi +V(\chi) \right)  \right],
\end{aligned}
\end{equation}
whose potential is
\begin{equation}
\label{eq:R3_potential}
\begin{aligned}
V(\chi)=\frac{\kappa_0}{72\alpha^2_0}e^{-2\chi}\left(1-\sqrt{1-4\alpha_0\left(1-e^\chi\right)}\right) \times \\ \times\left[-1+8\alpha_0\left(1-e^\chi\right)+\sqrt{1-4\alpha_0\left(1-e^\chi\right)}\right].
\end{aligned}
\end{equation}
In a way analogous to the Starobinsky model, the barred quantities are defined with respect to the conformal metric, and
\begin{equation}
\label{eq:Chi_function_R3}
\begin{aligned}
e^\chi=1+\frac{R}{\kappa_0}+\alpha_0\left(\frac{R}{\kappa_0}\right)^2.
\end{aligned}
\end{equation}

The next step is to obtain $V(\chi)$, $V'(\chi)$, and $V''(\chi)$ at leading order in the slow-roll approximation. Using the slow-roll parameter $\delta=e^{-\chi}$, we can approximate the potential and its derivatives as:\footnote{We are supposing that $\alpha_0\delta^{-1}\ll1$. See Ref.~\cite{Rodrigues2022} for details.}
\begin{align}
V(\chi) & \approx \frac{\kappa_0}{6}\left(1-2\delta-\frac{2\alpha_0}{3}\delta^{-1}\right), \label{eq:R3_slow-roll}\\
V'(\chi) & \approx \frac{\kappa_0}{3}\left(\delta-\frac{\alpha_0}{3}\delta^{-1}\right), \label{eq:dR3_slow-roll}\\
V''(\chi) & \approx -\frac{\kappa_0}{3}\left(\delta+\frac{\alpha_0}{3}\delta^{-1}\right).\label{eq:ddR3_slow-roll}
\end{align}
Substituting these results in Eqs. \eqref{eq:epsilon_slow-roll} and \eqref{eq:eta_slow-roll}, we obtain, at leading order in the slow-roll approximation
\begin{align}
\epsilon & \approx \frac{4}{3}\left(\delta-\frac{\alpha_0}{3}\delta^{-1}\right)^2, \label{eq:epsilon_R3}\\
\eta & \approx -\frac{8}{3}\left(\delta+\frac{\alpha_0}{3}\delta^{-1}\right).  \label{eq:eta_R3}
\end{align}
Once again, it is clear that for $\delta \ll1$ we have $\epsilon \ll1$ and $\eta \ll1$, and therefore the Universe is in an inflationary regime. 

The next step is to compute the number of e-folds $N_k$. Using Eqs. \eqref{eq:R3_slow-roll} and \eqref{eq:dR3_slow-roll} in \eqref{eq:Number-efolds_slow-roll}, we obtain
\begin{equation}
\label{eq:Number-efolds_R3}
\begin{aligned}
N_k \approx \frac{3}{8\delta_c}\ln\left[\frac{1+\sqrt{\frac{|\alpha_0|}{3\delta^2}}}{1-\sqrt{\frac{|\alpha_0|}{3\delta^2}}}\right].
\end{aligned}
\end{equation}
This last expression can be explicitly inverted, resulting in
\begin{equation}
\label{eq:Number-efolds_R3_both_case}
\begin{aligned}
\delta(N) =
\begin{cases}
\displaystyle
\sqrt{\frac{\alpha_0}{3}}
\left[
\frac{
\exp\!\left( \frac{8N_k}{3}\sqrt{\frac{\alpha_0}{3}} \right) + 1
}{
\exp\!\left( \frac{8N_k}{3}\sqrt{\frac{\alpha_0}{3}} \right) - 1
}
\right] \quad \text{for}
& \alpha_0 \ge 0, \\[1.2em]
\displaystyle
\sqrt{\frac{|\alpha_0|}{3}}
\left[\frac{1}{
\tan\!\left( \frac{4N_k}{3}\sqrt{\frac{|\alpha_0|}{3}} \right)
}\right] \quad \text{for}
& \alpha_0 < 0.
\end{cases}
\end{aligned}
\end{equation}
For small corrections to the Starobinsky model ($N_k\sqrt{|\alpha_0|}\ll1$), both branches reduce to a single expression
\begin{equation}
\label{eq:delta_Nk}
\begin{aligned}
\delta(N_k) \approx \frac{3}{4N_k}\left[1+\frac{1}{3}\left(\frac{4N_k}{3}\sqrt{\frac{|\alpha_0|}{3}}\right)^2\right].
\end{aligned}
\end{equation}
In the same way as for the Starobinsky model, we express the slow-roll parameters $\epsilon$ and $\eta$ in terms of the number of e-folds. By substituting Eq. \eqref{eq:delta_Nk} into Eqs. \eqref{eq:epsilon_R3} and \eqref{eq:eta_R3}, we obtain
\begin{align}
\epsilon & \approx \frac{3}{4N^2_k}\left[1-\frac{1}{12}\left(\frac{16N_k}{3}\sqrt{\frac{|\alpha_0|}{3}}\right)^2\right], \label{eq:epsilon_Nk_R3}\\
\eta & \approx -\frac{2}{N_k}\left[1+\frac{1}{12}\left(\frac{16N_k}{3}\sqrt{\frac{|\alpha_0|}{3}}\right)^2\right].  \label{eq:eta_Nk_R3}
\end{align}
With these results, we can relate them to observations using Eqs.~\eqref{eq:As_slow-roll}, \eqref{eq:ns_slow-roll}, and \eqref{eq:r_slow-roll}
\begin{align}
\kappa_0 & \approx 72\pi^2A_s\left(\frac{M_P}{N_k}\right)^2\left(1-\frac{\alpha_0}{3}\delta^{-2}\right)^2, \label{eq:kappa_R3}\\
n_s & \approx 1-\frac{2}{N_k}\left[1+\frac{1}{12}\left(\frac{16N_k}{3}\sqrt{\frac{|\alpha_0|}{3}}\right)^2\right], \label{eq:ns_R3}\\
r & \approx \frac{12}{N^2_k}\left[1-\frac{1}{12}\left(\frac{16N_k}{3}\sqrt{\frac{|\alpha_0|}{3}}\right)^2\right].  \label{eq:r_R3}
\end{align}
As can be seen from Eq.~\eqref{eq:ns_R3}, the spectral index $n_s$ depends on both $N_k$ and $\alpha_0$. In the next section, we show that even when $N_k$ is constrained within its theoretical range, it is still possible to choose $\alpha_0$ such that the $R^3$ model fits the ACT data remarkably well.

\section{\label{sec:III}Constraint on the e-folding number range $N_k$}

To derive a general constraint on $N_k$, it is necessary to connect the end of the inflationary period with the subsequent reheating phase and the later evolution of the Universe up to the present day. This requires constructing a cosmological scale that tracks the evolution of a given mode $k$ from the moment it exits the horizon during inflation until today. At the horizon exit, the mode satisfies the condition $k=a_kH_k$ where $a_kH_k$ represents the comoving Hubble radius at that instant. From this relation, we can write~\cite{Munoz2015, Mishra2021}
\begin{equation}
\label{eq:cosmological_scale}
\begin{aligned}
\frac{k}{a_0 H_k} = \left( \frac{a_k}{a_{end}} \right) \left( \frac{a_{end}}{a_{re}} \right) \left( \frac{a_{re}}{a_{eq}} \right) \left( \frac{a_{eq}}{a_0} \right),
\end{aligned}
\end{equation}
where $a_{end}$, $a_{re}$, $a_{eq}$, and $a_0$ are the values of the scale factor at the end of inflation, at the beginning of the radiation era, at the time of matter-radiation equality, and at the present day, respectively. Each fraction in the cosmological scale equation\footnote{We are neglecting the dark energy--dominated phase since it becomes relevant only very close to the present day.} represents a specific era of cosmic expansion, corresponding respectively to the periods between horizon crossing and the end of inflation, from the end of inflation to thermalization, from thermalization to matter-radiation equality, and from equality to the present day.

This cosmological scale is the most general one we can construct without introducing additional physics beyond the standard cosmological framework. This is because the end of cosmic inflation must smoothly connect the Universe to a decelerated phase accompanied by reheating, thereby preserving the homogeneity and isotropy of spacetime as well as the successful predictions of the standard model. A possible modification to these cosmological scales could arise if, between reheating and Big Bang Nucleosynthesis (BBN), the Universe underwent one or more phases of expansion different from a radiation-dominated era, i.e., with an equation of state parameter $p\neq \rho /3$. However, there is no fundamental motivation to postulate the existence of such an additional phase after reheating. Moreover, even from a purely phenomenological perspective, at high energies, the particles that dominate the energy density $\rho$ of the Universe are ultrarelativistic and weakly interacting. This holds true both to the particles of the Standard Model and to the vast majority of its extensions, naturally leading to a radiation-dominated evolution after reheating.

Taking the logarithm of Eq. \eqref{eq:cosmological_scale} resulting in
\begin{equation}
\label{eq:log_cosmological_scale}
\begin{aligned}
\ln\left( \frac{k}{a_0 H_k} \right) = -N_k - N_{re} + \ln\left( \frac{a_{re}}{a_{eq}} \right) + \ln\left( \frac{a_{eq}}{a_0} \right),
\end{aligned}
\end{equation}
where $N_{k}=\ln\left(\frac{a_{end}}{a_{k}}\right)$ and $N_{re}=\ln\left(\frac{a_{re}}{a_{end}}\right)$ with the latter being the number of e-folds during the reheating phase. The last two terms of Eq. \eqref{eq:log_cosmological_scale} can be rewritten making the standard assumption that entropy $S=ga^3T^3$ is conserved between the radiation and matter eras~\cite{Husdal2016, KolbTurner1990}. In this case, we can write
\begin{equation}
\label{eq:entropy_conserved}
\begin{aligned}
T_{eq} = \frac{a_0}{a_{eq}} T_0 \quad \text{and} \quad T_{eq} = \left( \frac{a_{re}}{a_{eq}} \right) \left( \frac{g_{re}}{g_0} \right)^{1/3} T_{re},
\end{aligned}
\end{equation}
where we consider that the relativistic degrees of freedom in the equivalence era and in the present day are identical $\left(g_0=g_{eq}\right)$. The next step is to work with the thermalization temperature $T_{re}$, reached at the beginning of the radiation era (the end of reheating). This temperature is related to the energy density at the end of reheating through the expression
\begin{equation}
\label{eq:energy_density}
\begin{aligned}
\rho_{re} = \frac{\pi^2}{30}g_{re}T^4_{re}.
\end{aligned}
\end{equation}
Using the covariant conservation equation
\begin{equation}
\label{eq:covariant_conservation_equation}
\begin{aligned}
\dot \rho =-3H\left(\rho+p\right),
\end{aligned}
\end{equation}
we can relate the energy density at the end of reheating $\rho_{re}$ to the energy density at the end of inflation $\rho_{end}$. If the equation of state of the cosmic fluid during reheating is known, this equation can be integrated to determine how the energy density evolves as a function of the scale factor. In practice, however, the reheating process is highly complex, involving nonperturbative and nonequilibrium phenomena such as inflaton decay, particle production, and thermalization, all of which depend on the detailed microphysics of the underlying model~\cite{Amin2014, Lozanov2019Lectures}. Nevertheless, it is possible to effectively describe the macroscopic behavior of this period using a phenomenological approach, where the dynamics are governed by an equation of state of the form
\begin{equation}
\label{eq:equation_of_state}
\begin{aligned}
p = \omega_{re}(N)\rho.
\end{aligned}
\end{equation}
Using Eqs. \eqref{eq:covariant_conservation_equation} and \eqref{eq:equation_of_state}, we get
\begin{equation}
\label{eq:energy_density_dot}
\begin{aligned}
\dot \rho =-3H\rho\left[1+\omega_{re}(N)\right].
\end{aligned}
\end{equation}
Since the equation of state parameter $\omega_{re}(N)$ may vary throughout the duration of reheating, it is convenient to define an effective constant value that encapsulates its overall behavior. This is done by introducing the average $\omega_{a}$, defined as
\begin{equation}
\label{eq:omega_a}
\begin{aligned}
\omega_a \equiv \frac{1}{N_{re}} \int_{0}^{N_{re}} \omega_{re}(N) \, dN \;\Rightarrow\; \rho_{re} = \rho_{end} e^{-3N_{re}(1+\omega_a)}.
\end{aligned}
\end{equation}
Given that the equation of state (EoS) during reheating evolves from values close to zero (typical of an inflaton-dominated universe) toward that of a radiation-dominated phase ($\omega \rightarrow1/3$), it is reasonable to assume that $\omega_{re}(N)$ behaves as a monotonically increasing function throughout reheating. This behavior is supported by numerical simulations, such as those involving lattice computations of inflaton fragmentation,\footnote{Lattice simulation–based numerical modeling suggests that the equation of state during reheating evolves monotonically~\cite{Maity2018, Lozanov2017, Lozanov2018}.} which confirm that the EoS parameter asymptotically approaches $1/3$ by the end of reheating. Under the assumption of monotonic increase, the average equation of state parameter $\omega_a$ must lie within the bounds
\begin{equation}
\label{eq:omega_a_limit}
\begin{aligned}
0<\omega_a<1/3.
\end{aligned}
\end{equation}
The closer $\omega_a$ is to $1/3$, the more efficient the reheating process becomes. By combining Eqs. \eqref{eq:energy_density} and \eqref{eq:omega_a} we obtain the reheating temperature in terms of $N_{re}$ and $\omega_a$
\begin{equation}
\label{eq:Tre1}
\begin{aligned}
T_{re} = \left( \frac{30 \rho_{end}}{g_{re} \pi^2} \right)^{1/4} 
\exp\!\left[ -\frac{3}{4} (1 + \omega_a) N_{re} \right].
\end{aligned}
\end{equation}
The next step is to obtain $N_{re}$ using Eqs. \eqref{eq:log_cosmological_scale}, \eqref{eq:entropy_conserved} and \eqref{eq:Tre1}:
\begin{equation}
\label{eq:Nre1}
\begin{aligned}
N_{re} = \frac{4}{3(\omega_a - \tfrac{1}{3})}
\left\{ N_k + \ln\!\left( \frac{\rho_{end}^{1/4}}{H_k} \right) + \ln\!\left( \frac{k}{a_0 T_0} \right) \right. \\
\left.
+ \frac{1}{4} \ln\!\left( \frac{30}{\pi^2} \right)
+ \frac{1}{12} \ln\!\left( \frac{g_{re}}{g_0^4} \right) \right\}.
\end{aligned}
\end{equation}

The first two terms depend explicitly on the inflationary model being considered. These quantities are determined by the form of the inflationary potential and by the dynamics of the inflaton field. The next two terms are numerical constants that arise from the standard thermodynamic and cosmological relations, such as the Stefan--Boltzmann law and entropy conservation. It is worth noting that, for the second logarithmic term, it is necessary to restore the physical units
\begin{equation}
\label{eq:physical_units}
\begin{aligned}
\log\!\left( \frac{k}{a_0 T_0} \right) = \ln\!\left( \frac{k c \hbar}{a_0 T_0 k_B} \right) = -65.08,
\end{aligned}
\end{equation}
where we consider the scale of interest $k=0.002{Mpc}^{-1}$ and $T_0=2.73K$. The last term depends on the particle content of the Universe during reheating and the present day, so this term is therefore sensitive to the underlying particle physics model and any beyond Standard Model extensions. Considering only the Standard Model fields as the matter content, we have $g_{re}=106.75$ and $g_0=3.94$~\cite{Husdal2016}. Thus,
\begin{equation}
\label{eq:gre/g0}
\begin{aligned}
\frac{1}{12} \ln\!\left( \frac{g_{re}}{g_0^4} \right) = -6.785 \times 10^{-2}.
\end{aligned}
\end{equation}
However, in practice, this term has a negligible numerical impact. The suppression arises from the overall logarithmic dependence together with the small prefactor $1/12$. Even in scenarios with a significantly larger value of $g_{re}$, such as supersymmetric extensions of the Standard Model, where $g_{re}$ can reach values $\mathcal{O}(10^{3})$~\cite{Abel1995}, the resulting contribution remains numerically small and, at most, shifts $N_k$ at the decimal level.

Using all these numerical results, we obtain:
\begin{equation}
\label{eq:Nre2}
\begin{aligned}
N_{re} = \frac{4}{3(\omega_a - \tfrac{1}{3})}
\left\{ N_k + \ln\!\left( \frac{\rho_{end}^{1/4}}{H_k} \right) - 64.87 \right\}.
\end{aligned}
\end{equation}
In canonical single-field models, $\rho_{end}$ can be written in terms of the potential $V(\chi)$. This follows from the condition that inflation ends when the slow-roll parameter reaches $\epsilon=1$. However, to proceed, it is necessary to specify the inflationary model.

\subsection{Reheating in Starobinsky and $R^3$ model}

The inflation ends when $\epsilon=1$. Substituting this condition into Eqs. \eqref{eq:First_FriedmannEquation}, \eqref{eq:Seccond_FriedmannEquation} and \eqref{eq:epsilon_definition} we get $V(\chi_{end})=\dot \chi^2_{end}$. Using this result in \eqref{eq:usual_notation}, we reach
\begin{equation}
\label{eq:rho_end}
\begin{aligned}
\rho_{end}=\frac{9M^2_P}{4}V(\chi_{end}).
\end{aligned}
\end{equation}
Now we can use Eq. \eqref{eq:Starobinsky_potential} and imposing $\epsilon=1$ in \eqref{eq:epsilon_slow-roll} to get an estimate for $\chi_{end}$
\begin{equation}
\label{eq:chi_end}
\begin{aligned}
\epsilon & \approx \frac{1}{3}\left[\frac{V'(\chi_{end})}{V(\chi_{end})}\right]^2=1 \Rightarrow e^{-\chi_{end}} \approx 2\sqrt{3}-3.
\end{aligned}
\end{equation}
Thus, substituting Eqs. \eqref{eq:chi_end} and \eqref{eq:Starobinsky_potential} into \eqref{eq:rho_end}, we obtain the energy density at the end of inflation
\begin{equation}
\label{eq:rho_end_Starobinsky}
\begin{aligned}
\rho_{end} \approx \frac{3\kappa_0}{2}\left(2-\sqrt{3}\right)^2M^2_P.
\end{aligned}
\end{equation}
To obtain $H_k$, we use the expression \eqref{eq:Vsta_slow-roll} in the equation \eqref{eq:First_FriedmannEquation}:
\begin{equation}
\label{eq:H_k}
\begin{aligned}
H_k \approx \frac{\kappa_0}{12}.
\end{aligned}
\end{equation}
Substituting Eqs. \eqref{eq:rho_end_Starobinsky}, \eqref{eq:H_k} and \eqref{eq:kappa_starobinsky} into Eqs. \eqref{eq:Nre2} and \eqref{eq:Tre1}, we get
\begin{equation}
\label{eq:Nre_starobinsky}
\begin{aligned}
N_{re} = \frac{4}{3(\omega_a - \tfrac{1}{3})}
\left\{N_k + \frac{1}{2}\ln\left(N_k\right)-\frac{1}{4}\ln\left(A_s\right) - 66 \right\},
\end{aligned}
\end{equation}
\begin{equation}
\label{eq:Tre_starobinsky}
\begin{aligned}
T_{re} = 1.02 \, \frac{\left(A_s\right)^{1/4}}{N_{k}^{1/2}} 
\exp\!\left[-\frac{3}{4}(1+\omega_{a})N_{re}\right] M_{Pl}.
\end{aligned}
\end{equation}
Using $A_{s} = 2.1 \times 10^{-9}$~\cite{Aghanim2021VI} in Eqs. \eqref{eq:Nre_starobinsky} and \eqref{eq:Tre_starobinsky}, we obtain
\begin{align}
N_{re} &= \frac{4}{3(\omega_a - \tfrac{1}{3})}
\left\{N_k + \frac{1}{2}\ln\left(N_k\right)-61 \right\}, \label{eq:Nre2_starobinsky}\\
T_{re} &= \frac{6.91\times 10^{-3}}{N_{k}^{1/2}} 
\exp\!\left[-\frac{3}{4}(1+\omega_{a})N_{re}\right] M_{Pl}. \label{eq:Tre2_starobinsky}
\end{align}

We now examine how constraints on $N_{re}$ and $T_{re}$ restrict the range of the number of e-folds $N_k$ of inflation. Let us first consider the case corresponding to the upper limit. This upper bound on $N_k$ follows from the physical requirement that $N_{re}\geq0$. In the most optimistic scenario, where reheating is extremely rapid and effectively instantaneous ($N_{re}=0$), we obtain the maximal possible value of $N_k$ compatible with this condition. According to Eq.~\eqref{eq:Nre2_starobinsky}, we then have
\begin{equation}
\label{eq:upper}
\begin{aligned}
N_k+\frac{1}{2}\ln(N_k)\le61 \Rightarrow N_k \le59.
\end{aligned}
\end{equation}
Note that this result is remarkably robust and largely independent of the specific particle physics model considered. As discussed earlier, even in scenarios where the effective number of relativistic degrees of freedom at reheating $g_{re}$ is significantly large, the impact leads to a small reduction in $N_{k}$.

We now turn to the lower limit of $N_{k}$. Determining this bound is less straightforward, as it requires specifying the minimal characteristics of the reheating phase to be considered. The post-inflationary epoch remains largely unconstrained by current particle physics knowledge. Although extensions of the Standard Model or potential nonminimal couplings might introduce additional interactions between the inflaton and matter fields, even in the absence of such mechanisms---or if their contribution to reheating is negligible---one can still adopt the minimal assumption that standard matter fields couple to gravity in the usual way. In the Einstein frame, this minimal coupling implies that the inflaton primarily decays into the Higgs doublet, with subdominant decay channels into gluon pairs. Based on these dominant processes, the inflaton decay rate can be expressed as~\cite{Gorbunov2013, Bernal2020}
\begin{equation}
\label{eq:decay_channel}
\begin{aligned}
\Gamma_{\phi} \approx \frac{1}{24\pi} \left[ (1 - 6\xi)^2 + \frac{49\alpha_{s}^{2}}{4\pi^{2}} \right] \frac{m_{\phi}^{3}}{M_{P}^{2}},
\end{aligned}
\end{equation}
where $\xi$ is the nonminimum coupling constant between the Higgs field and gravitation, that is $\xi|h|^{2}R$,\footnote{The nonminimal coupling is required to address the renormalizability issues of scalar fields in curved spacetime.~\cite{Callan1970}.} $\alpha_s$ is the coupling constant of QCD on the reheating energy scale\footnote{Extrapolating the experimental running of the strong coupling $\alpha_s$~\cite{dEnterria2015}, one can estimate that during reheating $0.01\lesssim\alpha_s\lesssim0.03$.} and $m_{\phi}$ is the effective mass of the inflaton, given by $m_{\phi}^2 \equiv V''(0)=\kappa_0/3$. 

Having determined the decay rate of the inflaton, we can estimate the reheating temperature $T_{re}$ as follows: the thermal equilibrium of the cosmic fluid is reached when $\Gamma_{\phi} \sim H$~\cite{Kofman1997}. Thus,
\begin{equation}
\label{eq:Tre_decay_channel}
\begin{aligned}
3M_{Pl}^{2}H_{re}^{2} = \rho_{re} = \frac{\pi^{2}}{30} g_{re} T_{re}^{4} \Rightarrow 
T_{re} \approx 0.5 \sqrt{\Gamma_{\phi} M_{Pl}},
\end{aligned}
\end{equation}
and using \eqref{eq:decay_channel}, we get
\begin{equation}
\label{eq:Tre3_starobinsky}
\begin{aligned}
T_{re} \approx 2.63 \times 10^{12} 
\left[\frac{1}{N_{k}}\right]^{\frac{3}{2}}
\sqrt{(1 - 6\xi)^{2} + \frac{49\alpha_{s}^{2}}{4\pi^{2}}} \ \text{GeV},
\end{aligned}
\end{equation}
where we use $g_{re}=106.75$ and $M_{P}=2.435\times10^{18}$GeV.

Although Eq~\eqref{eq:Tre3_starobinsky} depends on the value of $\xi$, the expression above provides a lower estimate for $T_{re}$. In fact, even though the Higgs field is conformally coupled to gravity, i.e., $\xi=1/6$, the inflaton decay channel into two gluons yields $T_{re}\sim10^8$GeV. Therefore, equation~\eqref{eq:Tre3_starobinsky} allows us to establish a minimum reheating temperature $T_{re}^{(min)}$ that is relatively insensitive to the details of the reheating phase. Even under minimal assumptions for reheating and with a conformal coupling of the Higgs field to gravity, we find

\begin{equation}
\label{eq:Tre_min}
\begin{aligned}
T_{re}^{(\text{min})} \approx 2.63 \times 10^{12}
\left[\frac{1}{N_{k}}\right]^{\frac{3}{2}}
\sqrt{\frac{49 \alpha_{s}^{2}}{4 \pi^{2}}} \, \text{GeV} \\
\Rightarrow
T_{re}^{(\text{min})} \approx 2.14 \times 10^{10}
\left[\frac{1}{N_{k}}\right]^{\frac{3}{2}} \text{GeV.}
\end{aligned}
\end{equation}
The lower limit for the number of e-folds $N_k$ is obtained by taking into account that $T_{re}\geq T_{re}^{(min)}$. Thus, by using Eqs. \eqref{eq:Tre_min}, \eqref{eq:Tre2_starobinsky} and \eqref{eq:Nre2_starobinsky}, we get
\begin{equation}
\label{eq:lower}
\begin{aligned}
N_{k} \exp \left\{
\frac{3(1 + \omega_{a})}{1 - 3 \omega_{a}}
\left[
N_{k} - \frac{1}{2} \ln\!\left(\frac{1}{N_{k}}\right) - 61
\right]
\right\}
\ge \times 10^{-6}.
\end{aligned}
\end{equation}
The decay speed of the exponential in Eq. \eqref{eq:lower} depends on the coefficient
\begin{equation}
\label{eq:wa_limit}
\begin{aligned}
3 < \frac{3\left(1+\omega_a\right)}{1-3\omega_a}<\infty,
\end{aligned}
\end{equation}
where the lower and upper limits were established from Eq.~\eqref{eq:omega_a_limit}. 

To construct a conservative scenario, we consider the case in which the thermalization process after inflation occurs as slowly as possible, minimizing the efficiency of energy transfer between the inflaton field and relativistic particles. This assumption corresponds to the slowest possible reheating phase and thus provides the most conservative estimate for the lower bound on $N_k$. Therefore, we will adopt $\omega_a \rightarrow0$ in Eq.~\eqref{eq:lower}:

\begin{equation}
\label{eq:lower_limit}
\begin{aligned}
N_{k} \exp \left\{3
\left[
N_{k} + \frac{1}{2} \ln\!\left(N_k\right) - 61
\right]
\right\}
\ge \times 10^{-6} \Rightarrow N_k \ge 53.
\end{aligned}
\end{equation}

Thus, we finally conclude that
\begin{equation}
\label{eq:range}
\begin{aligned}
53 < N_k < 59.
\end{aligned}
\end{equation}
This result represents the most general and model-independent bound that can be derived without introducing any exotic or non-standard physics beyond the minimal assumptions. The lower bound, though less straightforward to establish, is obtained under conservative conditions, assuming only gravitational couplings and minimal decay channels. Consequently, for well-motivated inflationary models such as Starobinsky, this range provides an excellent estimate. The analysis for the $R^3$ case follows an entirely analogous procedure to that of the Starobinsky model, and therefore, the complete derivation will not be repeated here. For readers interested in the detailed derivation within the $R^3$ framework, we refer to Ref.~\cite{Rodrigues2022}.\footnote{It is worth noting that a missing numerical factor of $1/2$ appears in Eq.~$(50)$ of reference~\cite{Rodrigues2022}; however, this omission does not affect the final results, as the factor is suppressed by the logarithmic dependence in the subsequent expressions.} 

\begin{figure}[b]
\includegraphics[width=0.5\textwidth]{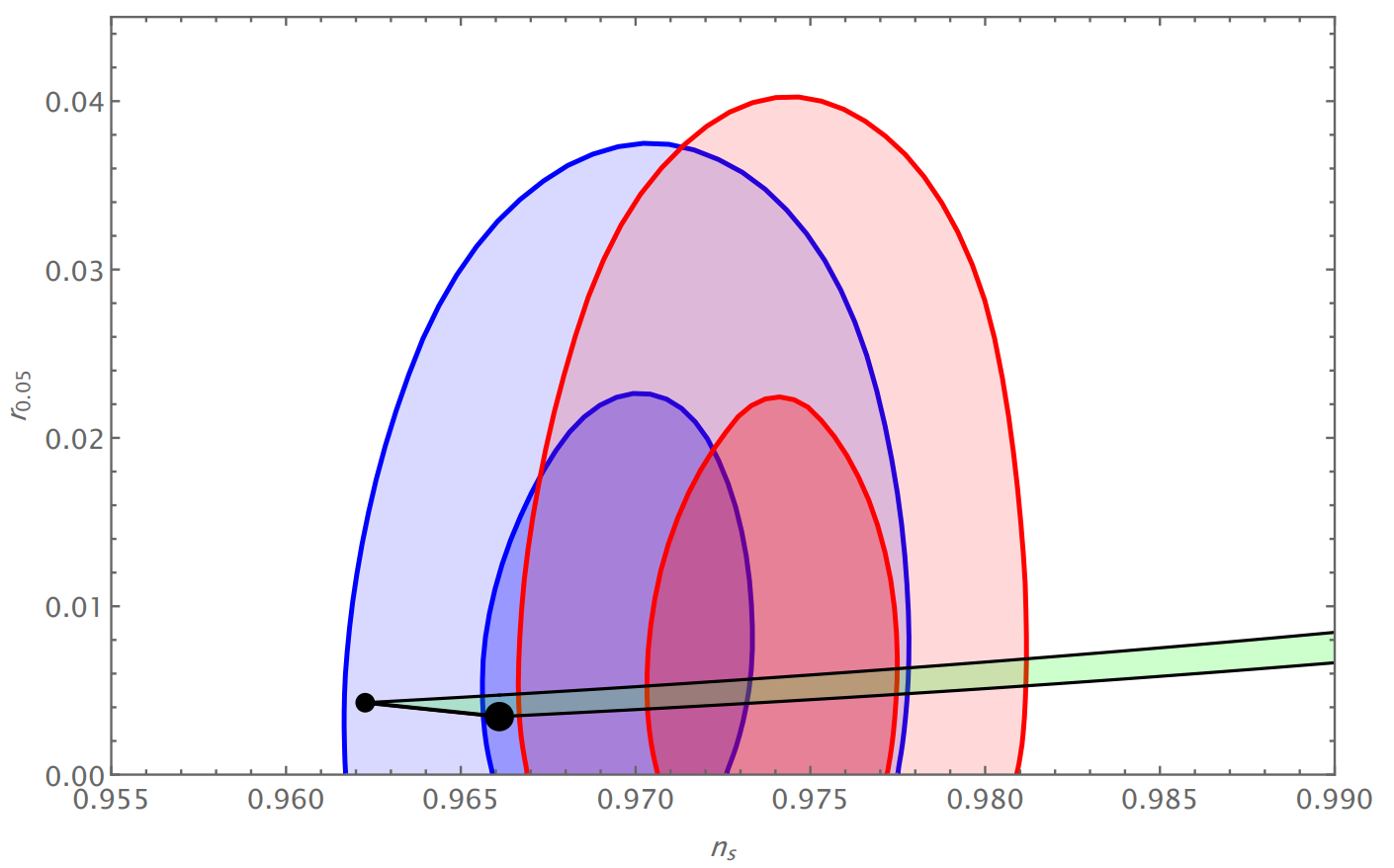}
\caption{Constraints on the scalar and tensor primordial power spectra for the Starobinsky and $R^3$ model. The constraints on $r$ are driven by the BK18 data, while the constraints on $n_s$ are driven by Planck (blue) or P-ACT (red). The combined dataset also includes CMB lensing and BAO in all cases. For more details about the dataset, see Ref.~\cite{ACTDR6b}. The light green region shows the theoretical evolution of the $R^3$ model. The black circles represent the Starobinsky model for $N_k=53$ (the smaller one) and $N_k=59$ (the bigger one). The darker contours represent the $68\%$ C.L., while the lighter ones correspond to the $95\%$ C.L..\label{fig:Constraint_Models}}
\end{figure}

With the constraint on $N_k$ obtained from the reheating analysis, we can now use the Planck and ACT data to test the consistency of both the Starobinsky model and its $R^3$ extension with observations. This can be achieved by using Eqs.~\eqref{eq:ns_starobinsky} and \eqref{eq:r_starobinsky} for the Starobinsky model, and Eqs.~\eqref{eq:ns_R3} and \eqref{eq:r_R3} for the $R^3$ extension, together with the range derived in Eq.~\eqref{eq:range}. Figure~\ref{fig:Constraint_Models}, shows the constraints on the scalar and tensor primordial power spectra at $k^\ast = 0.002\,\mathrm{Mpc}^{-1}$, in the $r-n_s$ parameter space.

The light green region shows the theoretical evolution of the $R^3$ model, while the black line connecting the black dots represents the Starobinsky solution. From Fig.~\ref{fig:Constraint_Models}, we see that within the interval $53 < N_k < 59$, the Starobinsky model is consistent with the Planck data~\cite{BICEPKeck2021}, but exhibits a tension greater than $2\sigma$ with the most recent P-ACT results~\cite{ACTDR6b}.

These results indicate that, once our theoretical constraint on $N_k$ is applied, the Starobinsky model becomes inconsistent with the latest CMB data. It should be noted, however, that this simplified approach merely tests the compatibility of the $\Lambda$CDM model with the inflationary phase. In the next section, we perform a complete statistical analysis by implementing the inflationary dynamics into the Boltzmann code CLASS. Under this approach, the Starobinsky model once again shows excellent agreement with the most recent observational data.

\section{\label{sec:IV}Implementation of Inflationary Dynamics in CLASS}

In this analysis, we modify the CLASS code version 3.3 to natively incorporate the Starobinsky and $R^3$ inflationary models. To capture subleading effects that can influence inflationary observables, we expand the potential as a Taylor series expansion up to fourth derivatives. Furthermore, rather than relying on the standard slow-roll approximation typically used in the literature to estimate the number of e-folds (see Eq.~\eqref{eq:Number-efolds_slow-roll}), we compute it by numerically integrating the full inflationary dynamics governed by Eqs.~\eqref{eq:First_FriedmannEquation}, \eqref{eq:Chi_FieldEquation}, \eqref{eq:epsilon_definition}, and \eqref{eq:Number-efolds_definition}. For readers interested in accessing the modified CLASS modules and the numerical scripts used in this work, please refer to the Data Availability section, where we provide the link to the public GitHub repository containing all implementation details.

Since we are implementing inflation directly into the CLASS code, the treatment of cosmological parameters differs substantially from the standard $\Lambda$CDM approach. In the conventional framework, the scalar amplitude $A_s$ and the scalar spectral index $n_s$ are treated as fundamental input parameters. In our implementation, however, $N_k$ and $\kappa_0$ become the primary parameters, while $A_s$ and $n_s$ are derived quantities.

In the standard scenario, six parameters are typically varied, $\left(\Omega_b,\Omega_{CDM},100\theta, \tau, A_s, n_s \right)$.
Here, since we focus exclusively on the inflationary sector, our discussion is restricted to the parameters related to the primordial power spectrum. For the Starobinsky model, the parameter set becomes $\left(\Omega_b,\Omega_{CDM},100\theta, \tau, \kappa_0, N_k \right)$, maintaining the same number of free parameters as in $\Lambda$CDM. For the $R^3$ model, an additional parameter $\alpha_0$ is introduced, yielding the set $\left(\Omega_b,\Omega_{CDM},100\theta, \tau, \kappa_0, N_k, \alpha_0 \right)$.

\subsection{Dataset and Priors}
To confront the predictions of the Starobinsky and \(R^3\) models with the most up-to-date observational evidence, we use:
\begin{itemize}
    \item \textbf{CMB data:} We combine large-scale data from Planck~\cite{Aghanim2021VI} with small-scale measurements from the Atacama Cosmology Telescope~\cite{ACTDR6a}, ensuring sensitivity to a large multipole range of the CMB spectrum.
    \item \textbf{Lensing:} Gravitational lensing measurements from both Planck~\cite{Aghanim2020VIII} and ACT~\cite{Qu2023} are included to help break parameter degeneracies and improve constraints on $A_s e^{-2\tau}$~\cite{Ade2013XVI}.
    \item \textbf{BAO:} Baryon acoustic oscillation data from the Dark Energy Spectroscopic Instrument (DESI DR2)~\cite{DESIDR2} are incorporated. BAO data help to break parameter degeneracies that limit analyses based solely on CMB observations~\cite{Ade2013XVI}.
\end{itemize}

The main goal of this work is to evaluate the validity of different inflationary scenarios — in particular, the Starobinsky and $R^3$ models — in light of the most recent observational data. To perform a consistent comparison, we employ the original Planck data combining information from both high- and low-multipole ranges in temperature and polarization. The lensing potential is taken exclusively from Planck measurements. In addition, we include BAO data from the DESI DR2 release, as these provide purely geometrical constraints that are largely unaffected by the complex systematic uncertainties present in other astrophysical probes (see Ref.~\cite{Ade2013XVI}).

The next dataset combination corresponds to the most recent observations currently available. We use CMB data from Planck and ACT DR6, along with the lensing reconstructions from both collaborations and BAO measurements from DESI DR2.
It is important to note that ground-based experiments such as ACT are not able to constrain the optical depth parameter $\tau$, as it is mainly sensitive to the low multipoles of the EE spectrum~\cite{Aghanim2020V}. In earlier analyses, this limitation was commonly addressed by imposing a Gaussian prior on $\tau$. However, with the latest SRoll2 likelihood, it is now possible to directly extract the large-scale polarization signal from the Planck HFI data using a refined mapmaking algorithm that minimizes temperature-to-polarization leakage and corrects for gain variations and other systematics~\cite{Delouis2019}. This approach provides a more robust, data-driven constraint on $\tau$, eliminating the need for an external prior.

We combine the aforementioned observations into two main dataset configurations used throughout this work:
\begin{itemize}
\item \textbf{P-LB:} This combination includes Planck temperature and polarization data, together with Planck lensing and BAO measurements from DESI DR2.
\item \textbf{P-ACT-LB:} This extended combination incorporates Planck and ACT DR6 CMB data, along with lensing reconstructions from both collaborations, BAO data from DESI DR2 and the SRoll2 likelihood.
\end{itemize}

The analysis is performed under the following set of priors show in table~\ref{tab:priors}.
\begin{table}
\caption{\label{tab:priors}Flat priors used in the Bayesian analysis.}
\begin{ruledtabular}
\begin{tabular}{lc}
\textrm{Parameter} & \textrm{Range} \\
\colrule
$\Omega_{b}h^2$ & $[0.001,\,1]$ \\
$\Omega_{c}h^2$ & $[0.001,\,1]$ \\
$100\,\theta$    & $[0.5,\,10]$ \\
$\tau$           & $[0.01,\,0.8]$ \\
$N_{k}$          & $[30,\,200]$ \\
$\kappa_0$       & $[1,\,100]\times10^{-14}$ \\
$\alpha_0$       & $[-3,\,3]\times10^{-10}$ \\
\end{tabular}
\end{ruledtabular}
\end{table}

For the Bayesian inference, we use Cobaya~\cite{Torrado2021} in its default configuration to explore the posteriors with a range of Monte Carlo samplers~\cite{Lewis2002, Lewis2013}. The convergence of the chains is monitored using the Gelman--Rubin criterion~\cite{Gelman1992}, following Cobaya’s standard convergence methodology. We use the GetDist package~\cite{Lewis2013GetDist} to analyze the samples and compute marginalized one- and two-dimensional posterior densities.

\subsection{Parameter Constraints}
The first quantitative results of our analysis based on the Starobinsky model are shown in table~\ref{tab:Starobinsky_Constraints}, which summarizes the constraints on the cosmological and primordial parameters for the P-LB and P-ACT-LB datasets. 
\begin{table}
\caption{\label{tab:Starobinsky_Constraints}
Posterior mean values and 65\% confidence intervals for the Starobinsky
model obtained using the datasets P-LB (Planck) and P-ACT-LB
(Planck+ACT). The parameters are organized into three categories:
inflationary, cosmological, and derived.}
\begin{ruledtabular}
\begin{tabular}{lcc}
\textrm{Parameter} & \textrm{P-LB} & \textrm{P-ACT-LB} \\
\colrule
$N_\mathrm{k}$ &
$61^{+4.5}_{-5.2}$ &
$63^{+5.0}_{-4.3}$ \\
$\kappa_0$ &
$\left(1.50^{+0.26}_{-0.23}\right)\times10^{-13}$ &
$\left(1.41^{+0.31}_{-0.26}\right)\times10^{-13}$ \\[2pt]
$100\,\theta_\mathrm{s}$ &
$1.04200 \pm 0.00027$ &
$1.04172 \pm 0.00029$ \\
$\Omega_\mathrm{b}h^2$ &
$0.02248 \pm 0.00013$ &
$0.02254 \pm 0.00012$ \\
$\Omega_\mathrm{c}h^2$ &
$0.1184 \pm 0.0007$ &
$0.1184 \pm 0.0006$ \\
$\tau_\mathrm{reio}$ &
$0.056 \pm 0.007$ &
$0.060 \pm 0.008$ \\[2pt]
$A_\mathrm{s}$ &
$\left(2.100^{+0.031}_{-0.030}\right)\times10^{-9}$ &
$\left(2.123^{+0.034}_{-0.032}\right)\times10^{-9}$ \\
$n_\mathrm{s}$ &
$0.9679^{+0.0024}_{-0.0028}$ &
$0.9692^{+0.0025}_{-0.0026}$ \\
$H_0$ &
$68.62 \pm 0.33$ &
$68.55 \pm 0.31$ \\
$\sigma_8$ &
$0.819 \pm 0.005$ &
$0.8235 \pm 0.006$ \\
\end{tabular}
\end{ruledtabular}
\end{table}

\begin{figure*}
\centering
\includegraphics[width=1\textwidth]{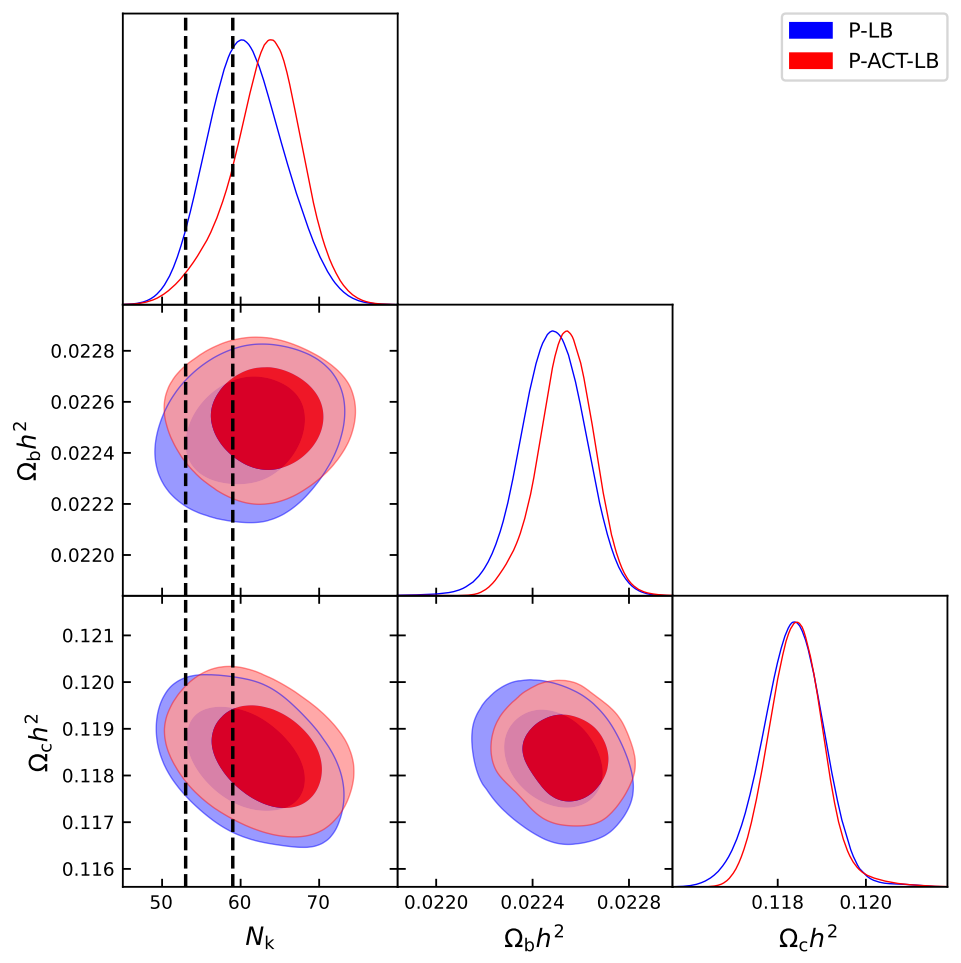}
\caption{Confidence regions for the Starobinsky model using P-LB (blue contours) and P-ACT-LB (red contours) datasets. The darker contours represent the $68\%$ confidence level ($1\sigma$), while the lighter ones correspond to the $95\%$ confidence level ($2\sigma$). The vertical dashed lines indicate the theoretical range of $N_k$ derived from the reheating analysis in Eq. \eqref{eq:range}.
\label{fig:Starobinsky_Full}}
\end{figure*}

The main result of our analysis is that the Starobinsky model remains fully consistent with the latest CMB observations. In the simplified approach — where the reheating range of $N_k$ is compared directly with the $\Lambda$CDM-derived spectral index — the model appears to be excluded at the $2\sigma$ level. However, this behavior disappears in the complete framework, as can be seen from the intersection between the shaded regions and the interval delimited by the dashed lines in Fig.~\ref{fig:Starobinsky_Full}. When the inflationary dynamics are implemented directly into the Boltzmann code, allowing for a self-consistent evolution of all cosmological parameters, the Starobinsky model once again shows excellent agreement with current data.

Nevertheless, it is important to emphasize the difference between the two datasets. When using Planck data alone, the allowed range of $N_k$ lies almost entirely within the $1\sigma$ region. In contrast, for the P-ACT dataset, only a small portion of this range falls within $2\sigma$, with most of it remaining consistent with current observations only at that confidence level. In any case, the Starobinsky model remains consistent with the latest CMB observations, although a mild tension appears to be emerging.

The main quantitative results of our analysis based on $R^3$ model are presented in table~\ref{tab:R3_Constraints}, which summarizes the constraints on the cosmological and primordial parameters for the P-LB and P-ACT-LB datasets. 
\begin{table}
\caption{\label{tab:R3_Constraints}
Posterior mean values and 65\% confidence intervals for the $R^3$ model
obtained using the datasets P-LB (Planck) and P-ACT-LB (Planck+ACT).
The parameters are organized into three categories: inflationary,
cosmological, and derived.}
\begin{ruledtabular}
\begin{tabular}{lcc}
\textrm{Parameter} & \textrm{P-LB} & \textrm{P-ACT-LB} \\
\colrule
$N_\mathrm{k}$ &
$58.0^{+2.2}_{-2.6}$ &
$57^{+6.8}_{-7.1}$ \\
$\kappa_0$ &
$\left(1.77^{+0.20}_{-0.19}\right)\times10^{-13}$ &
$\left(2.2^{+0.64}_{-0.63}\right)\times10^{-13}$ \\
$\alpha_0$ &
$\left(-2.6^{+2.5}_{-2.4}\right)\times10^{-5}$ &
$\left(-8.1^{+4.3}_{-4.7}\right)\times10^{-5}$ \\[2pt]
$100\,\theta_\mathrm{s}$ &
$1.04184 \pm 0.00037$ &
$1.04160 \pm 0.00034$ \\
$\Omega_\mathrm{b}h^2$ &
$0.02251 \pm 0.00010$ &
$0.02253 \pm 0.00012$ \\
$\Omega_\mathrm{c}h^2$ &
$0.1184 \pm 0.0006$ &
$0.1182 \pm 0.0007$ \\
$\tau_\mathrm{reio}$ &
$0.0562 \pm 0.013$ &
$0.061 \pm 0.014$ \\[2pt]
$A_\mathrm{s}$ &
$\left(2.110^{+0.028}_{-0.029}\right)\times10^{-9}$ &
$\left(2.123^{+0.033}_{-0.023}\right)\times10^{-9}$ \\
$n_\mathrm{s}$ &
$0.9692^{+0.0025}_{-0.0024}$ &
$0.9736^{+0.0033}_{-0.0031}$ \\
$H_0$ &
$68.58 \pm 0.27$ &
$68.58 \pm 0.38$ \\
$\sigma_8$ &
$0.821 \pm 0.006$ &
$0.824 \pm 0.008$ \\
\end{tabular}
\end{ruledtabular}
\end{table}

\begin{figure*}
\centering
\includegraphics[width=1\textwidth]{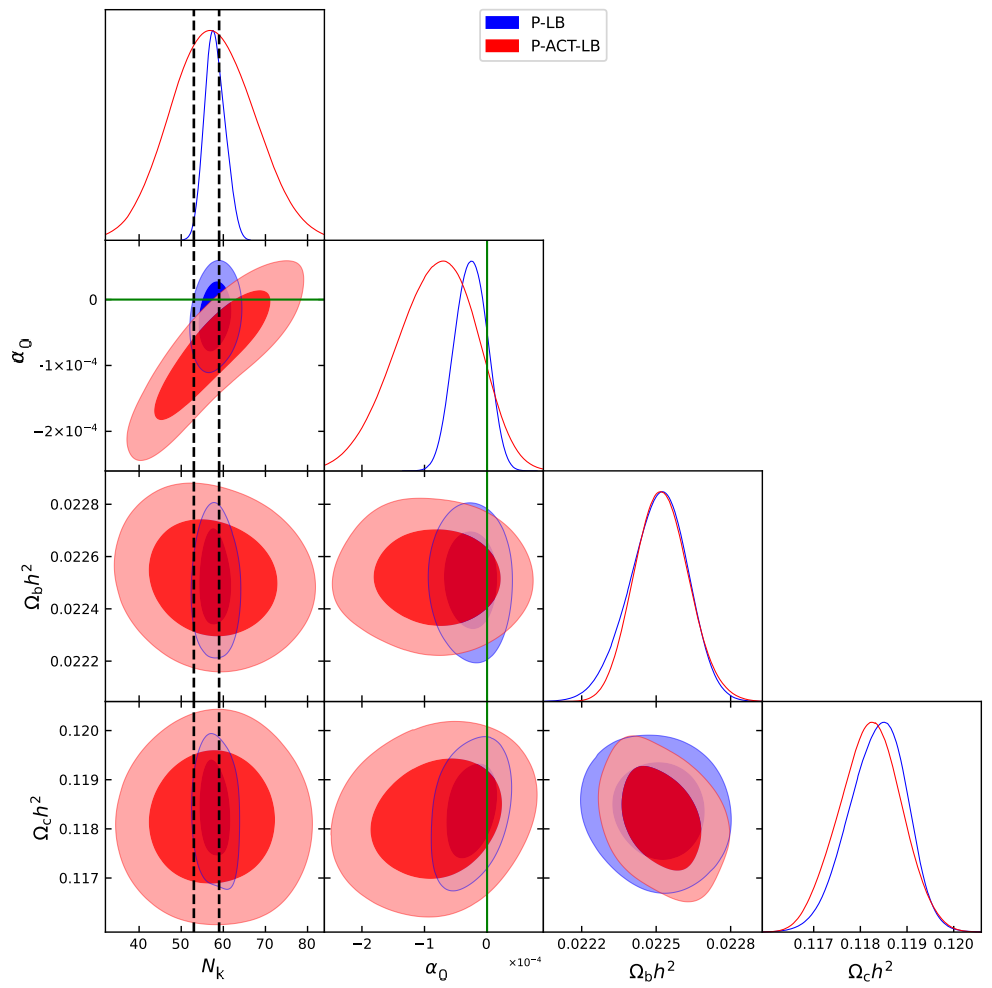}
\caption{Confidence regions for the $R^3$ model using P-LB (blue contours) and P-ACT-LB (red contours) datasets. The darker contours represent the $68\%$ confidence level ($1\sigma$), while the lighter ones correspond to the $95\%$ confidence level ($2\sigma$). The vertical dashed lines indicate the theoretical range of $N_k$ derived from the reheating analysis in Eq. \eqref{eq:range} and the green line indicates $\alpha=0$.
\label{fig:R3_Full}}
\end{figure*}

Figure~\ref{fig:R3_Full} shows the posterior distributions obtained from our Bayesian analysis. In this case, the peaks of the Gaussian posterior for $n_s$ lie well within the reheating-motivated range of the number of e-folds $N_k$. While the Starobinsky scenario exhibits a mild tension with the theoretically preferred window of $N_k$ when using the P-ACT dataset, the $R^3$ extension shifts the posterior for $N_k$ squarely into the expected interval for both the Planck and ACT data. The $R^3$ model introduces an additional degree of freedom that shifts $N_k$ in a direction preferred by the data, thereby improving the agreement with the reheating constraints.

Another important feature visible in Fig.~\ref{fig:R3_Full} concerns the joint posterior distribution of the parameters $N_k$ and $\alpha_0$. For the pure Planck data, no significant correlation is observed between these parameters. In the P-ACT-LB dataset, the Starobinsky limit ($\alpha_0=0$) remains consistent with current observations (see the green lines in Fig.~\ref{fig:R3_Full}); however, the inclusion of an additional degree of freedom in the extended $R^3$ model leads to a mild preference for a negative, nonzero value of $\alpha_0$. This suggests that the cubic correction may yield a slightly better fit to the combined datasets, indicating that the $R^3$ model is marginally favored by the data while still encompassing the Starobinsky case within its credible region.

\section{\label{sec:V}Conclusion}

In this work, we investigated how Starobinsky and $R^3$ inflationary models behave in light of the most recent observational data from the Atacama Cosmology Telescope, in comparison with the well-established Planck measurements. We first derived a theoretically motivated range for the number of e-folds $N_k$ by modeling the reheating phase under the minimal assumption of standard gravitational couplings between matter and geometry. Using this range, we confronted the models with the combined datasets Planck+BICEP3/Keck+BAO and Planck+ACT+BICEP3/Keck+BAO. In the simplified framework — where observables are computed in the slow-roll regime and compared with $\Lambda$CDM results — the Starobinsky model appears to be excluded at approximately the $2\sigma$ level when the ACT data are included.

In the second part of our analysis, we performed a more robust approach by implementing the inflationary dynamics directly into the Boltzmann code CLASS. This allowed us to self-consistently connect the inflationary phase to the late-time cosmological evolution. Within this framework, we found that the Starobinsky model remains fully consistent with the most recent CMB data. The theoretical range of $N_k$ lies well within the $1\sigma$ region for Planck data, while for the combined Planck+ACT dataset it extends slightly toward the $2\sigma$ region—indicating a mild tension.

For the extended $R^3$ model, the presence of one additional degree of freedom shifts the peaks of the Gaussian posterior for $n_s$ into the reheating-motivated window for the number of e-folds $N_k$, making the model fully consistent with both datasets. While the Planck+ACT combination shows a tendency toward a negative, nonzero value of $\alpha_0$, this should not be interpreted as evidence that the $R^3$ model is statistically preferred over the pure Starobinsky scenario, since the extended theory contains one extra parameter. What can be concluded is that both the original Starobinsky model and its $R^3$ extension remain fully compatible with current CMB observations.

It is important to emphasize that our results rely on a number of assumptions that must hold for the analysis to remain consistent. In particular, the theoretically motivated range for $N_k$ is valid only within the standard picture of a single, uninterrupted period of slow-roll inflation followed by a conventional reheating phase. If the early Universe dynamics were more exotic—for instance, if a second stage of inflation or an extended quasi–de Sitter phase occurred, or if additional dynamical fields became relevant during inflation—then the connection between inflationary scales and present-day observables would be modified, and our derived range for $N_k$ would no longer apply. In such scenarios, the constraints presented in this work could be substantially altered, and a dedicated analysis would be required.

Another important source of residual uncertainty arises from the treatment of astrophysical foregrounds and from the combination of datasets originating from different experiments. In this work, we combine Planck and ACT likelihoods, which rely on distinct instruments, observing strategies, frequency channels, and foreground–subtraction pipelines. While both collaborations provide state-of-the-art foreground modeling, small differences in component separation, beam systematics, and calibration procedures inevitably propagate into the inferred cosmological parameters. These residuals are especially relevant at high multipoles, where ACT dominates, and in the intermediate regime where both datasets overlap. Although the impact of these effects is mitigated by the use of publicly validated likelihoods, one must keep in mind that a full joint reanalysis with a unified foreground framework could lead to slight shifts in the posterior distributions. Our conclusions should therefore be interpreted within the context of these unavoidable experimental and modeling uncertainties.

Future CMB experiments, including the Simons Observatory (SO) and CMB-S4, together with large-scale structure measurements, are expected to play an important role in refining these constraints and testing the robustness of such extended inflationary models. The current uncertainty in the scalar spectral index from Planck is $\sigma(n_s)=0.040$~\cite{Aghanim2021VI}. The Simons Observatory aims to achieve $\sigma(n_s)=0.020$~\cite{Abitbol2025} by measuring primordial fluctuations over roughly twice the angular-scale range probed by Planck. Moreover, the combination of BAO and CMB-S4 observations is expected to extend this range by nearly a factor of four, potentially achieving $\sigma(n_s)\leq0.011$~\cite{Wu2014}.

In the framework adopted in this paper, where the number of e-folds $N_k$ is treated as a primary parameter and $n_s$ emerges as a derived quantity, the improvements in $n_s$ will translate into tighter and more accurate constraints on $N_k$, enabling a more precise mapping between inflationary dynamics and observable quantities. It is therefore likely that the enhanced sensitivity of SO and CMB-S4 will help determine whether small curvature corrections such as $R^3$ play a significant role in improving the concordance between inflationary models and forthcoming high–precision cosmological observations.

\begin{acknowledgments}
JBS acknowledges CNPq-Brazil for its support and LGM thank CNPq-Brazil for partial financial support—Grants: 307901/2022-0.
\end{acknowledgments}

\section*{Data Available}
As previously mentioned, this work employs observational data from the Planck\footnote{\href{https://pla.esac.esa.int/pla}{Planck Legacy Archive}}, 
ACT\footnote{\href{https://github.com/ACTCollaboration}{ACT Collaboration GitHub}}, 
DESI\footnote{\href{https://github.com/CobayaSampler/cobaya/tree/master/cobaya/likelihoods/bao/desi_dr2}{DESI likelihood (Cobaya)}} and BICEP/Keck\footnote{\href{http://bicepkeck.org/bk18_2021_release.html}{BICEP/Keck 2021 Release}}
 collaborations. All datasets are publicly available, and all relevant references for these datasets are cited throughout the paper. In addition, all modifications to the CLASS code used in this analysis are publicly available in our GitHub repository\footnote{\href{https://github.com/JeremiasBezerra/Starobinsky_R3-Inflation}{CLASS modifications}}.

\nocite{*}

\bibliography{apssamp}

\end{document}